\documentclass[12pt]{article}
\usepackage{epsfig}

\newcommand{\sect}[1]{\setcounter{equation}{0}\section{#1}}

\newcommand{\bea}{\begin{eqnarray}}
\newcommand{\eea}{\end{eqnarray}}

\def\be{\begin{equation}}
\def\ee{\end{equation}}
\def\ba{\begin{eqnarray}}
\def\ea{\end{eqnarray}}

\font\mybb=msbm10 at 11pt 
\def\bb#1{\hbox{\mybb#1}}

\def\bZ {\bb{Z}}
\def\bR {\bb{R}}

\newcommand{\no}{\noindent}

\def\CD{{\cal D}}
\def\CV{{\cal V}}
\def\CH{{\cal H}}
\def\CF{{\cal F}}
\def\CA{{\cal A}}
\def\CL{{\cal L}}
\def\CM{{\cal M}}
\def\CK{{\cal K}}
\def\CN{{\cal N}}
\def\CR{{\cal R}}

       \def\unit{\hbox to 3.3pt{\hskip1.3pt \vrule height 7pt width
.4pt
\hskip.7pt \vrule height 7.85pt width .4pt \kern-2.4pt \hrulefill
\kern-3pt \raise 4pt\hbox{\char'40}}}
\def\II{{\unit}}

\def\IK{\relax{\rm I\kern-.18em K}}

\def\IR{\relax{\rm I\kern-.18em R}}

\def\IZ{\relax\ifmmode\mathchoice {\hbox{\cmss Z\kern-.4em
Z}}{\hbox{\cmss Z\kern-.4em Z}} {\lower.9pt\hbox{\cmsss
Z\kern-.4em Z}} {\lower1.2pt\hbox{\cmsss Z\kern-.4em
Z}}\else{\cmss Z\kern-.4em Z}\fi}

\newcommand{\R}{{\bR}}
\newcommand{\Z}{{\bZ}}

\begin{document}

\baselineskip 18pt

\begin{titlepage}

\vfill
\begin{flushright}
\today\\
QMUL-PH-03-09\\
hep-th/0308133\\
\end{flushright}

\vfill

\begin{center}
{\bf \Large Compactifications with S-Duality Twists }

\vskip 10.mm {{Christopher M. Hull$^{1}$,}\\
and Aybike \c{C}atal-\"{O}zer $^{2}$}\\
\vskip 1cm

$^1${\it
Department of Physics\\
Queen Mary, University of London\\
Mile End Rd, London E1 4NS, UK
}\\

\vspace{2pt}

$^2${\it
Department of Physics\\
Middle East Technical University\\
\.{I}n\"{o}n\"{u} Bulvar\i Yolu, 06531 \\
Ankara, Turkey}\\

\vspace{6pt}

\end{center}
\par

\begin{abstract}
\noindent We consider generalised Scherk Schwarz reductions of
supergravity and superstring theories with    twists   by
electromagnetic dualities that are symmetries   of the   equations of
motion but
not of the action, such as the S-duality of $D=4, N=4$
super-Yang-Mills coupled to supergravity. The reduction cannot be
done on the action itself, but must be done either on the field
equations or on a duality invariant form of the action, such as
one in the doubled formalism in which potentials are introduced
for both electric and magnetic fields. The resulting theory in
odd-dimensions has massive form fields satisfying a self-duality
condition  $dA \sim m*A$. We construct such theories in $D=3,5,7$.

\end{abstract}
\vskip 1cm

\vfill \vskip 5mm \hrule width 5.cm \vskip 5mm {\small
\noindent $^1$ E-mail: c.m.hull@qmul.ac.uk \\
\noindent $^2$ E-mail: catal@metu.edu.tr }
\end{titlepage}


\sect{Introduction}

Twisted toroidal compactifications or Scherk-Schwarz reductions
are a useful way of introducing masses into supergravity and
string compactifications, generating a potential for the scalar
fields [1-19]. A theory in $D+1$ dimensions with global symmetry
$G$ can be compactified on a circle with fields not periodic but
with a $G$ monodromy around the circle, and the monodromy
introduces masses into the theory and breaks some of the symmetry.
The purpose here is to generalise such compactifications to the
case in which $G$ is a symmetry of the equations of motion only,
not of the action; we shall refer to such symmetries here as
S-dualities. A standard example is S-duality in 4-dimensions. The
heterotic string compactified to four dimensions has a classical
$SL(2,\R)$ symmetry which acts through electromagnetic duality
transformations and so is only a symmetry of the equations of
motion. In this case, we consider a circle reduction to three
dimensions with a monodromy in $SL(2,\R)$. In the quantum theory,
the  $SL(2,\R)$ symmetry is broken to $SL(2,\Z)$
$\cite{Sen:1994fa}$ and in that case the monodromy must be in
$SL(2,\Z)$ $\cite{chris2}$. We generalise this to other
dimensions, and discuss examples in $D=3,5$ and $7$ dimensions.

Consider a $D+1$ dimensional supergravity with a global symmetry
$G$. An element $g$ of the symmetry group acts on a generic field
$\psi$ as $\psi \to g[\psi]$. Consider now a dimensional reduction
of the theory to $D$ dimensions on a circle of radius $R$ with a
periodic coordinate $y\sim y+1$. In the twisted reduction, the
fields are not independent of the internal coordinate but are
chosen to have a specific dependence on the circle coordinate $y$
through the ansatz
\begin{equation}
\label{ansatz} \psi(x^{\mu}, y) = g(y) \, [ \psi (x^{\mu})]
\end{equation}
for some $y$-dependent group element $g(y)$ $\cite{chris2}$. An
important restriction on $g(y)$ is that the   reduced theory in
$D$ dimensions should be independent of $y$. This is achieved by
choosing
\begin{equation}
\label{massmatrix} g(y)= \exp (My)
\end{equation}
for some Lie-algebra element $M$. The map $g(y)$ is not periodic
around the circle, but has a {\it monodromy}
\begin{equation}
\label{monodromy}
       {\cal M} (g)= \exp M
\end{equation}

Many supergravity theories in $D+1=2n$ dimensions have a set of $n
$ form field strengths $ {H}_{n}^i$ where $i=1,...,r$ labels the
potentials, which typically satisfy a generalised self-duality
equation of the form
\begin{equation}
\label{duality} {H}_{n}^i= Q^i{}_j (\phi)  *  {H}_{n}^j
\end{equation}
where $Q^i{}_j$ is a matrix depending on the scalar fields $\phi$
and $  *$ is the Hodge dual in $D+1$ dimensions $\cite{julia}$.
For any $n$, consistency requires that $(Q^i{}_j (\phi) *)^2=1$,
so that if $(  *)^2=-1$, as in Lorentzian space of dimension $4m$,
then $Q^2=-\II$ and $Q$  is a complex structure, while if $(
*)^2=1$, as in Lorentzian space of dimension $4m+2$, then
$Q^2=\II$ and $Q$ is a product structure. In the theories we will
consider, the $ {H}_{n}^i$ transform in an $r$-dimensional
representation of a rigid duality group $G$. In $d=4$, $N=8$
supergravity, there are $r=56$ 2-form field strengths transforming
as a {\bf 56} of the duality group $G=E_7$
$\cite{cremmerjulia,cremmer}$. These split into 28 field strengths
$F=dA$ and 28 dual field strengths $\tilde F=\hat * F +\dots$,
with $Q$ a complex structure on $\R^{56}$. In $d=6$, $N=8$
supergravity, there are 5 3-form field strengths which split into
5 self-dual ones and 5 anti-self dual ones, and these 10 transform
as a {\bf 10} of $G=SO(5,5)$ $\cite{tanii}$. The 10 3-form field
strengths $\hat{H}_{n}^i$ with $i=1,...,10$, satisfy (anti)
self-duality constraints of the form $(\ref{duality})$ with $Q$
related to the $SO(5,5)$-invariant metric. In $d=8$ maximal
supergravity, there is a 3-form potential, and its field strength
and its dual combine into an $SL(2,\R)$ doublet, satisfying a
constraint of the form $(\ref{duality})$ with $Q = i \sigma_2$.

Our main interest here is in reductions in which the monodromy
${\cal M}\in G$ is a symmetry of the equations of motion but not
the action, acting on the field strengths $\hat{H}_{n}^i$ via
transformations involving Hodge or electromagnetic dualities, so that
they cannot be
realised locally on the fundamental  $n-1$ form potentials. We
find that (in the case in which $M$ is invertible) the field
strengths $\hat{H}_{n}^i$ satisfying the constraint
$(\ref{duality})$ give rise to $r$ {}  $n-1$ form potentials
$A_{n-1}^i$ in $2n-1$ dimensions satisfying   massive self-duality
constraints of the form
\begin{equation}
\label{eqn5} DA_{n-1}=\tilde M *A_{n-1}
\end{equation}
where $D$ is a gauge-covariant exterior derivative, $*$ is now the
Hodge dual in $D$ dimensions and the matrix  $\tilde{M}\propto
QM$. Such odd-dimensional self-duality conditions were first
considered in $\cite{Townsend:xs}$ and often occur in
odd-dimensional gauged supergravity theories, and follow from a
Chern-Simons  action with mass term of the form
\begin{equation}
\label{CS} L= P _{ij}  A^i \wedge D A^j + \hat M_{ij}A ^i \wedge *
A^j
\end{equation}
where $ \hat M =P \tilde M$ and $P _{ij}$ is a suitably chosen
constant matrix. In the general case in which $M$ is not
invertible, some of the gauge fields   remain massless.

In dimensionally reducing a theory with a twist that is a symmetry
of the equations of motion and not of the action, one needs to
reduce the equations of motion, not the action. However, for the
cases of interest here there is a doubled formalism $\cite{julia}$
in which dual potentials $\tilde A_{n-1}$ are introduced for each
$n-1$ form potential $ A_{n-1}$, in which the duality symmetry
becomes a symmetry of the action $S[A,\tilde A]$, which is
supplemented by a duality-invariant constraint that could be used
to eliminate $\tilde A$ in terms of $A$. This doubled action and
constraint can then be dimensionally reduced in the standard way
with a twist by the duality symmetry. This greatly simplifies the
calculations.

We apply these results to the reduction of supergravity theories
in $4,6,8$ dimensions, giving rise to supergravity theories in
$3,5,7$ dimensions with massive self-dual forms. This constructs
new supergravity theories in these dimensions and gives a
higher-dimensional origin for theories in $3,5,7$ dimensions with
Chern-Simons actions. In particular, for $D=3$, $A$ is a vector
field and this gives a higher dimensional origin for 3-dimensional
gauged supergravity theories, of the type discussed in
$\cite{nicolai}$ with Chern-Simons actions for some of the gauge
fields.

The plan of the paper is as follows. In section 2 we review the
Scherk-Schwarz mechanism, giving the  results for the twisted
reduction of  gravity coupled to scalars and  gauge potentials,
which are used in later sections.  We give a detailed analysis of the
general case in which the mass matrix is not invertible.
  In section 3 we review the
doubled formalism of \cite{julia}. In section 4 we perform a
twisted dimensional reduction in the doubled formalism, and hence
obtain the lagrangian for dimensional reductions with S-duality
twists. Finally, in  section 5,  we apply our results to the
reduction of supergravity theories in $4,6,8$ dimensions.

\section{Scherk Schwarz Reduction}

We will consider here Scherk-Schwarz dimensional reduction on a
circle from $D+1$ to $D$ dimensions, with a twist by an element of
a global symmetry $G$. The ansatz for dimensional reduction of a
generic field is $(\ref{ansatz})$ with $y$-dependence given by
$(\ref{massmatrix})$ with monodromy $\CM$ given by
$(\ref{monodromy})$ in terms of the mass-matrix $M$.  The mass
matrix $M$ introduces mass parameters into the theory, and fields
in non-trivial representations of the group $G$ typically become
massive with masses given in terms of $M$, or are ``eaten"  by
gauge fields that become massive in a generalised Higgs mechanism.
In particular, the scalar fields will obtain a scalar potential
given in terms of $M$. However, different mass-matrices can give
equivalent theories, and an important question is how to classify
the inequivalent theories. In $\cite{atish}$ it was shown that the
theories are determined by the monodromy ${\CM}$, not the mass
matrix $M$. Two reductions with different mass matrices $M,M'$ but
the same monodromy $\CM= e^M =e^{M'}$ give the same reduced
theory, provided the full spectrum of massive  states is kept, and
no truncation is made. In $\cite{chris2}$,   it was shown that
theories with monodromies in the same $G$ conjugacy class are
equivalent, so that the theories are classified by the  $G$
conjugacy classes. In   quantum string  theory, a global group of
the classical theory typically becomes a discrete gauge symmetry
      $G(\Z)$ $\cite{Hull:1994ys}$ and for such theories the monodromy
must be in  $G(\Z)$, giving   quantization conditions on the mass
parameters, and the distinct theories are determined by the
monodromy $\CM \in G(\Z)$ up to $G(\Z)$ conjugation. The  mass
matrix  $M$ generates a one dimensional subgroup $L$ of $G$, which
becomes a gauge symmetry of the reduced theory, so that such a
reduction of a supergravity gives a gauged supergravity
$\cite{chris3,bergshoeff,Meessen:1998qm,atish}$.

Our main interest will be in the reduction of supergravity and
superstring theories. Extended supergravity theories typically
have a global symmetry $G$ and the scalars take values in the
coset space $G/H$ where $G$ is a non-compact group  and $H$ is the
maximal compact subgroup of $G$. The scalar sector of the theory
is then invariant under the group $G$ and this symmetry typically
extends to the full theory for supergravities in odd dimensions.
In some even dimensional theories, the symmetry $G$ extends to a
symmetry of the equations of motion only, acting through duality
transformations exchanging field equations with Bianchi
identitites.

The theory can be formulated with a local $H$ symmetry as well as
a global $G$ symmetry. The scalars in the coset space $G/H$ can be
represented by a vielbein $\CV(x) \in G$ which transforms under
global $G$ and local $H$ transformations as
\begin{eqnarray}\label{eqn7}
\CV & \rightarrow & h(x) \CV, \ \ \ \ \ \ \ h(x) \in H \nonumber \\
\CV & \rightarrow & \CV \ g,  \ \ \ \ \ \ \ \ \ \ g \in G
\end{eqnarray}
The lagrangian is
\begin{equation}\label{eqn8}
         L =- \frac{1}{2} {\rm tr}[d {\cal{ V}} \CV ^{-1} \wedge *
d{\cal{V}} \CV ^{-1}].
\end{equation}
In this formulation there are an extra $dim (H)$ non-physical
scalars which can be gauged away using the local $H$ symmetry.
Here $\CV,g,h$ can be taken to be matrices in some representation
of $G$. We will present our results for real representations of
$G$ such that the representatives of $H$ are orthogonal matrices
$h^T h=\II$ so that $\delta _{ab}$ is an invariant, but the
generalisation to other representations is straightforward.

An alternative formulation that does not involve extra scalars is
to use a metric $\CK$ on $G/H$ instead of a vielbein, transforming
as (for a real representation of $G$)
\begin{equation}
\label{eqn9}  \CK\to g^T \CK g
\end{equation}
Such a metric can be constructed from the vielbein as $\CK_{ij} =
\delta_{ab} \CV_{\ i}^{a} \CV_{\ j}^{b} $, where $i$ and $a$ are
the curved and flat indices respectively. $\CK$ is invariant under
local $H$ transformations as $h^T h= \II$.  This means that the
non-physical scalars drop out in this formulation, without any
need for gauge-fixing. (For complex representations with
$h^\dagger h=\II$, we would use the hermitian metric $\CK =
\CV^{\dagger} \CV$ transforming as $\CK\to g ^{\dagger} \CK g$.)
The lagrangian can be written in terms of $\CK$ as
\begin{equation}\label{eqn10}
         L = \frac{1}{4} {\rm tr}[d {\cal{K}}^{-1} \wedge *
d{\cal{K}}].
\end{equation}

An example which will play a central role in what follows is a
theory of gravity coupled to scalars in the coset $G/H$ and a set
of $r$ {} $n-1$ form  gauge potentials $A_{n-1}^i$ with $n$-form
field strengths $H_n^i=dA_{n-1}^i$ (where $i=1,...,r$ )
transforming in  a real  $r$-dimensional representation of the
symmetry group $G$. We take $\CV$ to be an $r\times r$ matrix
acting in the $r$-dimensional representation of $G$ and consider
the theory in $D+1$ dimensions and work with the metric
$\CK_{ij}$. The lagrangian is
\begin{equation}\label{actionhigh}
\mathcal{L} =  R * 1 + \frac{1}{4} {\rm tr}(d \CK \wedge *
d\CK^{-1})  - \frac{1}{2} H_{n}^{T} \CK  \wedge * H_{n}
\end{equation}
     The action is invariant under the rigid $G$ symmetry
\begin{equation}
\label{transform} \delta A\to L ^{-1}A, \qquad \delta \CK\to L ^T
\CK L
\end{equation}
where $L ^i{}_j$ is a $G$-transformation in the {\bf r}
representation, and the spacetime metric is invariant.  In later
sections, we will be particularly interested in the case in which
$D+1=2n$, but for now we will keep $D,n$ arbitrary.

For example, in the case $G=SL(2,\R)$, $H=SO(2)$, there are two
scalars in the theory, which we will denote $\phi$ and $\chi$,
which parametrise the scalar coset $SL(2,\R)/SO(2)$. The matrix
$\CV$ (in  the doublet representation of $SL(2,\R)$) is a general
$SL(2,\R)$ matrix, which can be given, in terms of
       $\phi$ and $\chi$ and a non-physical scalar $\theta$ that
parameterises the $SO(2)$ subgroup, by
\begin{equation}
\mathcal{V} = h e^{\phi/2} \left(
\begin{array}{cc}\label{vielbein}
e^{- \phi} & 0 \\
-\chi & 1
\end{array} \right)
\end{equation}
where $h$ is an   $SO(2) $ matrix
\begin{equation}
h=  \left( \begin{array}{cc}\label{orthogonal}
\cos \theta & \sin \theta \\
- \sin \theta & \cos \theta
\end{array} \right)
\end{equation}
Then
\begin{equation}
\CK = e^{\phi} \left( \begin{array}{cc}\label{matrix}
e^{-2 \phi} + \chi^{2} & -\chi \\
-\chi & 1
\end{array} \right).
\end{equation}
and  the lagrangian (\ref{actionhigh}) can be written as
\begin{equation}
\CL = R * 1 - \frac{1}{2} d \phi \wedge * d \phi - \frac{1}{2}
e^{2 \phi} d \chi \wedge * d \chi -\frac{1}{2} (e^{-\phi} + e^\phi
\chi^2)H^1 \wedge * H^1 - \frac{1}{2} e^\phi H^2 \wedge * H^2 -
\chi e^\phi H^1 \wedge * H^2
\end{equation}
and is independent of $\theta$.

We now reduce the lagrangian $(\ref{actionhigh})$ on a circle with
a twist given by a monodromy $\CM =e^M \in G$ with the ansatz
$(\ref{ansatz})$. For the remainder of this section, we
distinguish  the $D +1$-dimensional fields from $D$-dimensional
ones by a hat. The metric is invariant under the global symmetry
group so we use the standard Kaluza-Klein ansatz
\begin{equation}\label{metric}
        d\hat{s}^{2} = e^{2 \alpha \varphi} ds^{2} + e^{2 \beta
        \varphi} (dy + \CA)^{2}
\end{equation}
     so that  the Einstein-Hilbert term in $(\ref{actionhigh})$
reduces to
\begin{equation}\label{einstein}
{\cal{L}}_{g} = R * 1 - \frac{1}{2} d\varphi \wedge * d\varphi -
e^{-2(D-1)\alpha \varphi} \frac{1}{2} \CF \wedge * \CF.
\end{equation}
Here $\varphi$ is the scalar field coming from the reduction of
the metric, $\CF = d\CA$ and $\CA$ is the graviphoton. The
constants $\alpha$ and $\beta$ depend on $D$ and are:
\begin{equation}\label{alfa}
\alpha^{2} = \frac{1}{2(D-1)(D-2)}, \ \ \ \ \ \beta = -(D-2)
\alpha.
\end{equation}
    From $(\ref{ansatz})$ and $(\ref{transform})$ the
ansatz for the scalar fields and the 3-form fields is
\begin{eqnarray}
\hat{\CK}(x, y) & = &
\lambda^{T}(y) \CK(x) \lambda(y) \label{ansatzscalar}\\
\hat{A}_{n-1}(x, y) & = & \lambda^{-1}(y) [A_{n-1}(x) + A_{n-2}(x)
\wedge dy]. \label{ansatzA}
\end{eqnarray}
      where $\lambda(y) = e^{M y}$.

By using the ansatz $(\ref{ansatzscalar})$  one finds that the
reduction of the scalar kinetic term \\ $\frac{1}{4} {\rm tr}(d
\hat{\CK} \wedge * d\hat{\CK}^{-1})$ from  $D+1$ dimensions to
$D$ dimensions gives a scalar kinetic term  plus a scalar potential
$\cite{SS}$:
\begin{equation}\label{scalar1}
      \CL_{s}  =   \frac{1}{4} {\rm tr}(\CD \CK \wedge * \CD
\CK^{-1}) + V(\phi)
\end{equation}
      where
      \begin{eqnarray}\label{scalar2}
\CD \CK & = & d \CK - (M^{T} \CK + \CK M) \wedge \CA \\
\CD \CK^{-1} & = & d \CK^{-1} + (M \CK^{-1} + \CK^{-1} M^{T})
\wedge \CA \nonumber
\end{eqnarray}
and the scalar potential $V(\phi)$ is
\begin{equation}\label{scalar3}
      V(\phi) =    - \frac{1}{2}  e^{2(D-1) \alpha
\varphi} {\rm tr}(M^{2} + M \CK^{-1} M^{T} \CK) * 1
\end{equation}
The ansatz $(\ref{ansatzA})$ implies
\begin{equation}\label{field1}
\hat{H}_{n}(x, y) = e^{-My} H_{n}(x) + e^{-My} H_{n-1}(x)\wedge
(dy + \mathcal{A})
\end{equation}
      for the $n$-form field strengths $\hat{H}_{n} = d
\hat{A}_{n-1}$.
Here the $D$-dimensional
      field strengths are
\begin{equation}\label{field2}
H_{n-1}(x) = dA_{n-2} - (-1)^{n-1} M A_{n-1}, \ \ \ \ H_{n}(x) =
dA_{n-1} - H_{n-1} \wedge \CA .
\end{equation}
      Reduction of the  kinetic term   gives
\begin{equation}\label{field3}
\hat{H}^{T}_{n} \hat{\CK} \wedge \hat{*} \hat{H}_{n}
      \rightarrow [e^{-2(n-1) \alpha \varphi} H^{T}_{n} \CK \wedge *
H_{n} + e^{2(D-n) \alpha \varphi} H^{T}_{n-1} \CK \wedge *
H_{n-1}] \wedge dy
\end{equation}

Collecting   the results  we can now write down the
$D$-dimensional lagrangian as:
\begin{equation}\label{lagrangian}
\CL_{D}  =  \CL_{g} + \CL_{b} + \CL_{s}
\end{equation}
     where
\begin{eqnarray}
\CL_{g} & = &  R * 1 - \frac{1}{2} d\varphi \wedge * d\varphi -
\frac{1}{2} e^{-2(D-1) \alpha \varphi}
\CF_{2} \wedge * \CF_{2}  \label{auxiliary} \\
\CL_{s} & =  &  \frac{1}{4} {\rm tr}(\CD\CK \wedge * \CD\CK^{-1})
- \frac{1}{2}  e^{2(D-1) \alpha \varphi} {\rm tr}(M^{2} + M
\CK^{-1} M^{T} \CK) * 1 \nonumber
\end{eqnarray}
and
\begin{equation}\label{auxiliaryy}
\CL_{b}  =  -\frac{1}{2} e^{-2(n-1) \alpha \varphi} H^{T}_{n}
\CK \wedge * H_{n}  -\frac{1}{2} e^{2(D-n) \alpha
\varphi} H^{T}_{n-1} \CK \wedge * H_{n-1} \nonumber \\
\end{equation}

The  field strengths  $(\ref{field2})$   are invariant under the
following   gauge transformations:
\begin{equation}\label{gauge}
         \delta A_{n-1} = d \Lambda, \ \ \ \ \ \ \ \delta A_{n-2} =
(-1)^{n-1} M \Lambda .
\end{equation}
If $M$ is invertible, these can be used to gauge $ A_{n-2}$ to
zero by performing the gauge transformation:
\begin{equation}\label{gaugetransformation}
A_{n-1} \rightarrow A_{n-1} + (-1)^{n-1} M^{-1} d A_{n-2}.
\end{equation}
     In this gauge the $D$-dimensional field strengths become
\begin{equation}\label{fieldstrength1}
H_{n}  =  D A_{n-1} = dA_{n-1} - (-1)^{n} M A_{n-1} \wedge
\mathcal{A}
\end{equation}
\begin{equation}\label{fieldstrength2}
H_{n-1}  =  (-1)^{n} M A_{n-1}.
\end{equation}
Then $A_{n-2}$ disappears from the theory, and the term $H_{n-1}
\wedge *H_{n-1} $ is a mass term for $A_{n-1}$. The degrees of
freedom represented by the $r$ fields $A_{n-2}$ have been absorbed
by the $r$
       $(n-1)$-form fields $A_{n-1}$ which have
become massive. Now $H_{n}=  D A_{n-1}$ is a gauge covariant
derivative where the gauge group is the subgroup of $G$ generated
by $M$ and the corresponding gauge field is the graviphoton $\CA$.

Now we will analyze the case $M$ is not invertible. It is useful
to work with flat indices $H^{a} = \CV_{\ i}^{a} H^{i}$, $A^{a} =
\CV_{\ i}^{a} A^{i}$. Then
$H^{a} = \CD A^{a} = dA^{a} + \omega^{a}{}_{ b} A^{b}$ where
$\omega$ is the connection 1-form $\omega^{a}_{\ b} = \CV_{\
i}^{a} (d\CV^{-1})^{i}_{\ b}$. The groups $G$ arising in the
supergravity theories of interest here all have a $G$-invariant
matrix $\Omega$ which is symmetric if $n$ is odd and
anti-symmetric if $n$ is even
\begin{equation}
\label{omega22} \Omega^{ab}=(-1)^{n-1}\Omega^{ba}.
\end{equation}
Using this,  we   introduce $\bar{H}_{a} = (\Omega^{-1})_{ab}
H^{b}$ and $M^{ab} = M^{a}{}_{ c} \Omega^{cb}$.
    Now one has
\begin{equation}\label{deneme}
H_{n-1}^{a} = \CD A_{n-2}^{a} - (-1)^{n-1} M^{ab} \bar{A}_{(n-1)
b} , \ \ \ \ \ \bar{H}_{(n) a} = \CD \bar{A}_{(n-1) a} -
\bar{H}_{(n-1) a} \wedge \CA = \tilde \CD \bar{A}_{(n-1) a}
\end{equation}
where $ \tilde \CD$ is the covariant derivative with connections $
\omega^{a}{}_{ b} $ and $ \CA $.

Note that $\CM = e^{M}$ and $\CM ^{T}\Omega ^{-1}\CM =
\Omega^{-1}$ since $\CM \in G$ and $\Omega $ is $G$-invariant.
(For complex representations, the condition is $\CM^{\dagger}
\Omega^{-1} \CM = \Omega^{-1}$.) As a result the mass matrix $M^a{}_b
$
satisfies:
\begin{equation}\label{inf}
         M^{T} \Omega ^{-1}+ \Omega^{-1} M = 0.
\end{equation}
  From $(\ref{omega22})$ and $(\ref{inf})$ it follows that $M^{ab}$ is
a
symmetric matrix if $n$ is even and antisymmetric if $n$ is odd:
\begin{equation}
\label{M2} M^{ab}=(-1)^{n}M^{ba}.
\end{equation}
Let the dimension of $ker(M)$ be $l$. Now the matrix $M^{ab}$ can
be brought into the canonical form
\begin{equation}\label{canonical}
M^{ab} = \left(\begin{array}{cc} 0 & 0 \\
                                    0 & m^{\alpha' \beta'}
                                    \end{array}\right)
\end{equation}
where $m^{\alpha' \beta'}$ is an invertible $(r-l) \times (r-l)$
matrix which is diagonal if $n$ is even and skew-diagonal if $n$
is odd. Here we have split the indices $a \rightarrow (\alpha,
\alpha')$ where $\alpha$ runs from 1 to $l$ and $\alpha'$ runs
from $l+1$ to $r$. Similarly the gauge fields $A$ can be written
in the block form
\begin{equation}\label{block}
A = \left( \begin{array}{c}
                  A^{\alpha}  \\
                  A^{\alpha '}                \end{array} \right)
\end{equation}
Performing the gauge transformation
\begin{equation}\label{deneme2}
       \bar{A}_{(n-1) \alpha'} \rightarrow \bar{A}_{(n-1) \alpha'} +
(-1)^{n-1}
       (m^{-1})_{\alpha' \beta'} \CD A_{n-2}^{\beta'}
\end{equation}
    one sees that the $r-l$ fields $
\bar{A}_{(n-1) \alpha'}$ become massive, having eaten the $r-l$
fields $A_{n-2}^{\alpha'}$, while    $ A_{n-2}^{\alpha}$ and $
\bar{A}_{(n-1) \alpha}$ both remain in the theory as massless
gauge fields, with $l$ of each. The field strengths for the
($n-2)$-form fields in $(\ref{deneme})$ become
\begin{equation}\label{deneme4}
       H_{n-1}^{\alpha'} = (-1)^{n} m^{\alpha' \beta'} \bar{A}_{(n-1)
       \beta'}, \ \ \ \ \ H_{n-1}^{\alpha} = \CD A_{n-2}^{\alpha}
\end{equation}
    and hence the   term
$(\ref{auxiliaryy})$ can be written as
\begin{eqnarray}\label{deneme3}
\CL_{b} & = & -\frac{1}{2}e^{-2(n-1) \alpha \varphi}
\delta^{\alpha \beta} \bar{H}_{(n) \alpha} \wedge * \bar{H}_{(n)
\beta} -\frac{1}{2}e^{-2(n-1) \alpha \varphi} \delta^{\alpha'
\beta'}
\bar{H}_{(n) \alpha'}\wedge * \bar{H}_{(n) \beta'}  \\
& & - \frac{1}{2} e^{2(D-n) \alpha \varphi} \delta_{\alpha \beta}
\CD A_{n-2}^{\alpha}\wedge * \CD A_{n-2}^{\beta} - \frac{1}{2}
e^{2(D-n) \alpha \varphi} (m^{T}m)^{\alpha' \beta'} \bar{A}_{(n-1)
\alpha'} \wedge *\bar{A}_{(n-1) \beta'}. \nonumber
\end{eqnarray}
We have chosen the   normalisation of $\Omega$ so that
$\Omega ^{ac} \Omega ^{bd} \delta _{cd} = \delta ^{ab}$.

    The gauge group, the
couplings and the scalar potential of the $D$-dimensional theory
found above are given explicitly in terms of the mass matrix $M$,
and two theories are distinct if the monodromies are in distinct
$G$-conjugacy classes.
       For the case $G = SL(2, \R)$ there are three conjugacy classes,
the hyperbolic, elliptic and parabolic conjugacy classes and so
there are three distinct reductions $\cite{chris2}$. The
hyperbolic, elliptic and parabolic monodromy matrices  and mass
matrices can be taken to be:
\begin{equation}\label{sl2mono}
\CM_{h} = \left( \begin{array}{cc}
                    e^{m} & 0 \\
                   0 & e^{-m}
                     \end{array} \right), \ \ \ \ \ \CM_{e} = \left(
\begin{array}{cc}
                    \cos m & \sin m \\
                   -\sin m & \cos m
                     \end{array} \right), \ \ \ \ \ \CM_{p} = \left(
\begin{array}{cc}
                    1  & m \\
                   0 & 1
                     \end{array} \right).
\end{equation}
\begin{equation}\label{sl2mass}
M_{h} = \left( \begin{array}{cc}
                    m & 0 \\
                   0 & -m
                     \end{array} \right), \ \ \ \ \ M_{e} = \left(
\begin{array}{cc}
                    0 & m \\
                   -m & 0
                     \end{array} \right), \ \ \ \ \ M_{p} = \left(
\begin{array}{cc}
                    0  & m \\
                   0 & 0
                     \end{array} \right).
\end{equation}
The mass  matrix generates a one-parameter subgroup of $SL(2,
\IR)$ and this subgroup will be the gauge group in the lower
dimensional theory. Thus compactification with $M_{e}$ will give a
compact gauging $SO(2)$ whereas compactification with $M_{h}$ and
$M_{p}$ will give rise to $SO(1,1)$-gauged lower dimensional
theories $\cite{chris3}$. (Note that in the special case in which
$n=3$, there will be extra vector gauge fields in $D$ dimensions
from the reduction of the 2-form gauge fields, and strictly
speaking the gauge group is
     $ISO(2)$, $ISO(1,1)$ or the Heisenberg group for the elliptic,
hyperbolic and parabolic cases, respectively $\cite{atish}$.)

The parabolic mass matrix $M_{p}$ is not invertible, and has a
one-dimensional kernel, i.e. $r=2,l=1$, so that $\alpha $ and
$\alpha '$ both take only one value and $A^a =(A^1, A^{1'})$. In
this case the matrix $m^{\alpha' \beta'}$ in $(\ref{canonical})$
is the $1 \times 1$ matrix $(-m)$  and  from
     the gauge transformation $(\ref{deneme2})$
      it can be seen that  the
$(n-1)$-form  $\bar A_{n-1\ 1'}$ eats   the $(n-2)$-form
$A_{n-2}^{1'}$ and becomes massive. The remaining $n-1 $ form
$\bar A_{n-1\ 1}$ and $n-2$ form $A_{n-2}^{1}$ gauge fields remain
massless.


\section{The Doubled Formalism}

Typically a $D=2n$ dimensional supergravity theory has a global
symmetry group $G$ which can be realised at the level of field
equations but not the action, as $G$ acts on $n$-form field
strengths $H=dA$  through   electric-magnetic duality
transformations. In such cases it is possible to construct a
manifestly $G$-invariant lagrangian   that depends on the
potentials $A$ and
      dual potentials $\tilde A$.
The dual fields  are   regarded as independent fields, but the
field equations are supplemented with
     a $G$-covariant constraint relating the $n$-form
field strengths $d\tilde A$ to $dA$, keeping the number of
independent degrees of freedom correct. The new lagrangian is
equivalent to the original one as the two  yield equivalent field
equations when the constraint is taken into account.

In this section we will review this formalism, which was
introduced in $\cite{julia}$ where it was called the  \lq doubled
formalism'. We will first consider the case $G=SL(2, \IR)$ and
then give   the general case in the following subsection.

\subsection{G = SL(2, $\IR$) Case}

Consider the following lagrangian in $ 2n$ dimensions with $n$
even
\begin{equation}\label{lag1}
\mathcal{L} = - \frac{1}{2} d \phi \wedge * d \phi - \frac{1}{2}
e^{2 \phi} d \chi \wedge * d \chi - \frac{1}{2} e^{-\phi} F_{n}
\wedge * F_{n} - \frac{1}{2} \chi F_{n} \wedge F_{n}
\end{equation}
\noindent Here  $F_{n} = dA_{n-1}$ and $\phi$ and $\chi$ are
scalar fields.  The field equations of this lagrangian have an
$SL(2, \IR)$ S-duality invariance (for even $n$) acting on $F$
through electromagenetic duality transformations, as we now
discuss.

Defining a new $n$-form $G$ by
\begin{equation}\label{G}
G_{n} =   \ \frac{\delta \mathcal{L}}{\delta F_{n}} = -e^{-\phi}
* F - \chi F
\end{equation}
\noindent the lagrangian ($\ref{lag1}$) can be written as
\begin{equation}\label{lag2}
\mathcal{L} = \frac{1}{4} {\rm tr}(d \CK \wedge * d \CK^{-1})
+\frac{1}{2} F \wedge  G
\end{equation}
\no where $\CK$ is as in $(\ref{matrix})$. The Bianchi identity
and the equation of motion for the $n$-form field strength $F_{n}$
are
\begin{eqnarray}\label{fe1}
dF_{n} & = & 0 \nonumber \\
dG_{n} = d(-e^{-\phi} * F_{n} - \chi F_{n}) & = & 0
\end{eqnarray}
\noindent which can be  combined as
\begin{equation}\label{fe2}
d\mathcal{H}_{n} = 0
\end{equation}
\noindent where $\mathcal{H}_{n}$ is the $SL(2, \IR)$ doublet
\begin{equation}\label{doublet1}
\mathcal{H}_{n} = \left( \begin{array}{c}
F_{n}    \\
G_{n}
\end{array} \right).
\end{equation}

The field equations are manifestly $SL(2,\R)$ invariant, but the
$F \wedge G$ term in the lagrangian $(\ref{lag2})$ is not
invariant. However, an invariant lagrangian can be constructed as
in $\cite{julia}$ if the field equation $dG_n=0$ is solved by
introducing a    dual potential $\tilde{A}_{n}$ so that
$G_n=d\tilde{A}_{n}$, which can be combined with $A_n$ to form an
$SL(2,\R)$ doublet, with field strengths $H_n^{i}$ given by
\begin{equation}\label{doublet2}
H_{n} = \left( \begin{array}{c}
dA_{n}    \\
d\tilde{A}_{n}
\end{array} \right).
\end{equation}
Then the natural $SL(2, \IR)$ invariant lagrangian is
\begin{equation}\label{doubled}
\mathcal{L}' =  \frac{1}{4} {\rm tr}(d \CK \wedge * d\CK^{-1}) -
\frac{1}{4} H_{n}^{i} \CK_{ij}\wedge * H_{n}^{j}.
\end{equation}
\no which is of the form considered in the previous section.

For this action, both $A_{n-1}$ and $\tilde A_{n-1}$ are
independent fields, so that the number of $n-1$ form degrees of
freedom has been doubled. To halve them again,  for even $n$ this
action can be supplemented by the $SL(2,\R)$ covariant constraint
$\cite{julia}$
\begin{equation}\label{constraint1}
H_{n}^{i} = J^{i}_{\ j} * H_{n}^{j}
\end{equation}
\noindent where $J$ is the $SL(2, \IR)$ matrix
\begin{equation}\label{J}
J^{i}_{\ j}= \Omega ^{ik}\CK_{kj}.
\end{equation}
\noindent Here $\Omega$ is the $SL(2, \IR)$ invariant matrix
\begin{equation}
\Omega = \left( \begin{array}{cc}\label{Omega}
0 & 1 \\
-1 & 0
\end{array} \right).
\end{equation}
\no Note that the matrix $J$ in $(\ref{J})$ satisfies $J^{2} = -
\II$, so that this constraint is consistent in $2n$ dimensions
with even $n$ in which $(*)^2=-1$. The general  case including odd
$n$ will be discussed in the next subsection.

The field equations derived from ($\ref{doubled}$) are
supplemented by the extra condition ($\ref{constraint1}$). Then
the field equations derived from ($\ref{doubled}$), together with
the constraint ($\ref{constraint1}$) which can be used to rewrite
all terms involving $\tilde A$ in terms of $A$, gives precisely
the field equations derived from the original action, so that the
original lagrangian ($\ref{lag1}$) and the $SL(2,\R)$ invariant
lagrangian ($\ref{doubled}$) and constraint ($\ref{constraint1}$)
are equivalent.
Note that the conventional normalisation of the gauge
field kinetic term in  (\ref{doubled}) has a factor of $1/4$ in the
doubled formalism
whereas in equation (\ref{actionhigh}) it had a factor of $1/2$.
Similar factors of $1/4$ will occur in the  normalisations of kinetic
terms in subsequent
lagrangians in the doubled formalism.

\subsection{The General Formalism}
The doubled formalism of the last section can be generalised
$\cite{julia}$. Consider the following lagrangian in $2n$
dimensions
\begin{equation}\label{lagr1}
\CL = - \frac{1}{2}R_{IJ}F_{n}^I \wedge * F_{n} ^J-
\frac{1}{2}S_{IJ} F_{n} ^I\wedge F_{n}^J+ {\rm L}(\Phi)
\end{equation}
where $F^{I}_{n} = dA^{I}_{n-1}$ with $I=1,...,k$  are   $k$ field
strengths and $\Phi$ denotes all the remaining fields, including
scalar fields.  The matrices $R_{IJ},S_{IJ}$ are functions of the
scalar fields and they satisfy $R_{IJ} = R_{JI}$ and $S_{IJ} =
(-1)^{n-1} S_{JI}$. It is useful to define $G^{I}_{n}$ as
\begin{equation}\label{GG}
G^{I}_{n} =  \ \frac{\delta \mathcal{L}}{\delta F^{I}_{n}}.
\end{equation}
\no so that the lagrangian can be written as
\begin{equation}\label{lagr2}
\CL = \frac{1}{2} F^{I}_{n} \wedge G^{I}_{n} + {\rm L}(\Phi)
\end{equation}

The field equations and Bianchi identities can be combined as
\begin{equation}\label{fe3}
d\mathcal{H}_{n} = 0
\end{equation}
\noindent where $\mathcal{H}_{n}$ is
\begin{equation}\label{doublet3}
\mathcal{H}_{n} = \left( \begin{array}{c}
F_{n}  ^I  \\
G_{n}^I
\end{array} \right).
\end{equation}
with    $r=2k$ components.

Such systems arise in supergravity theories, and typically the
field equations and Bianchi identities have a global symmetry $G$
under which  $\CH_{n} $ transforms as a $2k$-dimensional
representation of $G$, and $ {\rm L}(\Phi) $ is $G$-invariant. The
group $G$
     has a constant  invariant matrix
$\Omega^{ij}=\Omega^{ab}(\CV^{-1})^i{}_a (\CV^{-1})^j{}_b$, where
$i,j=1,...,2k$ are indices for the {\bf 2k} representation of $G$,
satisfying
    $ \Omega^{ij}=(-1)^{n-1}\Omega^{ji}$.

As before we introduce   potential fields $\tilde{A}^{I}_{n-1}$
with $G_n^I=d\tilde A_{n-1}^I$ to form
\begin{equation}\label{doublet4}
H_{n} = \left( \begin{array}{c}
dA^{I}_{n}    \\
d\tilde{A}^{I}_{n}
\end{array} \right)
\end{equation}
  transforming in the  {\bf 2k} representation of $G$. Then the
system can be described by the $G$-invariant lagrangian
\begin{equation}\label{doubled2}
\CL' = -\frac{1}{4} H_{n}^{T} \CK \wedge * H_{n} + {\rm L}(\Phi),
\end{equation}
  together with a  constraint
\begin{equation}\label{constraint2}
H_{n} =Q  * H_{n}.
\end{equation}
  where $Q^i{}_j$ is a $2k\times 2k $ matrix given in terms of the
scalar fields by
\begin{equation}
\label{Q} Q^i{}_j=\Omega^{ik} \CK_{kj}
\end{equation}
  Here $\CK_{ij}$ is given in terms of $R_{IJ},S_{IJ}$ by
\begin{equation}
\label{KK} \CK = \left(\begin{array}{cc}
                           R + SR^{-1}S^{T} & -SR^{-1} \\
                           -R^{-1}S^{T} \ \ \ \ & \ R^{-1}
                           \end{array}\right)
\end{equation}
In the supergravity applications we will be considering, the
scalars take values in a coset $G/H$ and $\CK_{ij}$ is the
symmetric matrix representing the scalar fields, as described in
section 2. \no Note that
\begin{equation}\label{QQ}
       Q^2=(\Omega \CK)^{2} = (-1)^{n-1} \II.
\end{equation}
  so that the constraint ($\ref{constraint2}$) is consistent as
for $2n$-dimensional Lorentzian space-time $* * H_{n} = (-1)^{n-1}
H_{n}$. It was shown in $\cite{julia}$ that the field equations
from $(\ref{doubled2})$ are equivalent to those from
($\ref{lagr1}$) together with the constraint
($\ref{constraint2}$).


\section{Reduction with  Duality Twist  }

The  theory with lagrangian   ($\ref{lagr1}$)   has a global
symmetry $G$ of the equations of motion which acts via duality
transformations. In this section we will dimensionally reduce on a
circle from $D+1=2n$ to $D$ dimensions with a twist that has
monodromy $\CM$ in $G$. For some choices of monodromy $\CM$ in
$G$, this is in fact a symmetry of the action and this is a
standard Scherk-Schwarz reduction, as in section 2. If it is only
a symmetry of the equations of motion, then we use the doubled
formalism of section 3 with lagrangian ($\ref{doubled2}$)
supplemented by the constraint ($\ref{constraint2}$). The
lagrangian ($\ref{doubled2}$) is of the same form as
($\ref{actionhigh}$), so the Scherk-Schwarz reduction of the
action proceeds as in section 2. This is supplemented by the
constraints arising from the dimensional reduction of
($\ref{constraint2}$). The field equations in $2n-1$ dimensions
are then those from the reduced action together with the reduced
constraints, and we go on to seek an action in $2n-1$ dimensions
that gives both the constraints and the reduced field equations.

\subsection{Dimensional Reduction in the Doubled Formalism}

The lagrangian ($\ref{doubled2}$) in the doubled formalism is of
the same form as ($\ref{actionhigh}$), but with an extra factor of
$1/2$ in the normalisation of  the gauge field kinetic term. The
Scherk-Schwarz reduction of the lagrangian  ($\ref{actionhigh}$)
    was already discussed in section 2, where we showed that it yields
the lagrangian  $(\ref{lagrangian})$ in $D$ dimensions. It follows
that the reduction of ($\ref{doubled2}$) should give
$(\ref{lagrangian})$ but now with   $(\ref{auxiliaryy})$ divided
by two to give:
\begin{equation}
   \CL_{b}  =  -\frac{1}{4} e^{-2(n-1) \alpha \varphi} H^{T}_{n}
\CK \wedge * H_{n}  -\frac{1}{4} e^{2(D-n) \alpha
\varphi} H^{T}_{n-1} \CK \wedge * H_{n-1}   \label{quarter}
\end{equation}
Just as the lagrangian ($\ref{doubled2}$) should be supplemented
by the $D+1$ dimensional constraint ($\ref{constraint2})$  in
order to give the correct $D+1$ dimensional field equations, the
$D$ dimensional lagrangian $(\ref{lagrangian})$ with
($\ref{auxiliary})$, ($\ref{quarter})$ should be supplemented by
the constraint which is   obtained by the dimensional reduction of
($\ref{constraint2})$. In this section we will describe the
reduction of the $D+1$-dimensional constraint
$(\ref{constraint2})$. Note that it is $G$-covariant, so the $y$
dependence of the fields in the ansatz $(\ref{ansatz})$  cancels
out in the reduction.

Using the ansatz $(\ref{ansatzscalar})$, $(\ref{ansatzA})$ the
$D+1$ dimensional constraint $(\ref{constraint2})$ reduces to the
$D$-dimensional constraint:
\begin{equation}
H_{n}  =  e^{\gamma} Q * H_{n-1} \label{constraintlow}
\end{equation}
\no where $Q$ is as in $(\ref{Q})$, $\CK$ is given by $(\ref{KK})$
and we have defined $\gamma \equiv 2(D-n) \alpha \varphi$. As a
result,
       the $n$-form field
strengths are dual to the $n-1$-form field strengths. The
constraint $(\ref{constraintlow})$ can be rewritten using flat
indices as
\begin{equation}\label{flat}
\bar{H}_{(n) a} = e^{\gamma} \delta_{a b} * (\CD A_{(n-2)}^{b} +
(-1)^{n} M^{bc} \bar{A}_{(n-1) c}).
\end{equation}
     For an untwisted reduction (i.e. one with $M=0$, so that it is  a
standard reduction) this constraint can be used to eliminate the
$2k$ potentials $A_{n-1}$ so that the theory can be written in
terms of the $2k$ potentials $A_{n-2}$ (or alternatively the
potentials $A_{n-2}$ can be eliminated and the theory written in
terms of the $A_{n-1}$, or more generally in terms of $s$
potentials $A_{n-2}$  and $2k-s$ potentials $A_{n-1}$). In the
twisted case with invertible $M$,
      one can go to the gauge in which the fields
$A_{n-2}^{i}$  are set zero, as was discussed in section 2. In
this gauge the field strengths $H_{n}$ and $H_{n-1}$ are given in
$(\ref{fieldstrength1})$ and $(\ref{fieldstrength2})$ so that the
duality condition $(\ref{constraintlow})$ is:
\begin{equation}
D { A}_{n-1}  = (-1)^{n} e^{\gamma} \tilde{M} * {A}_{n-1}
\label{constraint3}
\end{equation}
where $\tilde{M} = Q M$. This is a massive self-duality condition
for the $2k$ potentials ${ A}_{n-1}$. Such self-duality conditions
in odd dimensions were introduced in $\cite{Townsend:xs}$. The
self-duality constraint $(\ref{constraint3})$ implies the massive
field equation (suppressing non-linear terms)
\begin{equation}
*D*D { A}_{n-1}  =   e^{2 \gamma} \tilde{M}^2
      {A}_{n-1}+\dots  \label{mass3}
\end{equation}
with mass matrix proportional to $\tilde{M}^2$. However, the
constraint $(\ref{constraint3})$ halves the number of degrees of
freedom of a massive $n-1$ form field.

It is instructive to  check the number of physical degrees of
freedom. In $d$ dimensions a massless $p$ form gauge field $A_{p}$
has $c^{d-2}_p$ degrees of freedom, where $c^{s}_p$ is the binomial
coefficient
$$
c^{s}_p=\frac{(s)!}{p!(s-p)!}
$$
while a massive $p$ form gauge field has $c^{d-1}_p$ degrees of
freedom. The $k$ gauge fields $A_{n-1}^I$ in $2n$ dimensions have
$kc^{2n-2}_{n-1}$ degrees of freedom, which can be represented by
the $2k $ gauge fields $A_{n-1}^i$ (with  $2kc^{2n-2}_{n-1}$
degrees of freedom) together with $k$ constraints
($\ref{constraint2})$ that halve the number of degrees of freedom
again. In an untwisted reduction, each massless $n-1$ form gauge
field in $2n$ dimensions gives rise to a massless $n-1$ form gauge
field and a massless $n-2$ form gauge field, and the number of
degrees of freedom is correct as
$$c^{2n-2}_{n-1}=c^{2n-3}_{n-1}+c^{2n-3}_{n-2}$$
However, the number of degrees of freedom of a massive $p$-form in
$d-1$ dimensions is $c^{d-2}_p$, which is the same as the number
of degrees of freedom of a massless $p$-form in $d$ dimensions,
and in the twisted reduction with invertible $M$, all the $n-1$
forms in $2n$ dimensions give rise to massive $n-1$ forms in
$2n-1$ dimensions. We have $2k$ massive gauge fields $A_{n-1}^i$
in $2n-1$ dimensions which have $2kc^{2n-2}_{n-1}$ degrees of
freedom, but the self-duality constraints $(\ref{constraint3})$
remove half of the degrees of freedom, leaving $kc^{2n-2}_{n-1}$
degrees of freedom, as required.

    When $M$ is not invertible, the field strengths are given by
($\ref{deneme4})$. Then the constraint $(\ref{flat})$ takes the
form (dropping the coupling to the graviphoton again)
\begin{eqnarray}\label{deneme5}
      \CD \bar{A}_{(n-1) \alpha'} &=& (-1)^{n} e^{\gamma}
\delta_{\alpha' \beta'}
       * m^{\beta' \gamma'} \bar{A}_{(n-1) \gamma'} \\
\CD \bar{A}_{(n-1) \alpha} &=& e^{\gamma} \delta_{\alpha \beta} *
\CD A_{n-2}^{\beta} \label{deneme6}
\end{eqnarray}
Before imposing the constraint, the $r-l$ fields $ \bar{A}_{(n-1)
\alpha'}$ are massive, having eaten the $r-l$ fields
     $A_{n-2}^{\alpha'}$,
while    $ A_{n-2}^{\alpha}$ and $ \bar{A}_{(n-1) \alpha}$ both
remain in the theory as massless gauge fields, with $l$ of each,
as was seen in section 2. So before imposing the constraint the
total number of degrees of freedom is $$(r-l)c^{2n-2}_{n-1} + l
c^{2n-3}_{n-2} + l c^{2n-3}_{n-1} = r c^{2n-2}_{n-1} = 2k
c^{2n-2}_{n-1}.$$
    Imposing the constraint imposes
self-duality on the massive fields $ \bar{A}_{(n-1) \alpha'}$,
halving the number of degrees of freedom, and relates $
\bar{A}_{(n-1) \alpha}$ to $ A^{\alpha}_{n-2}$, so that half of
them can be eliminated (e.g.  $ A^{\alpha}_{n-2}$ can be
eliminated leaving $ \bar{A}_{(n-1) \alpha}$, or $ \bar{A}_{(n-1)
\alpha}$ can be eliminated leaving  $ A^{\alpha}_{n-2}$). Thus one
is left with $k c^{2n-2}_{n-1}$ degrees of freedom, as required.

The field equations from the $D$-dimensional lagrangian
$(\ref{lagrangian})$ with $(\ref{auxiliary})$, $(\ref{quarter})$
are  supplemented by the $D$ dimensional constraint
$(\ref{constraintlow})$. This implies that  the field strengths
$H_{n}$ and $H_{n-1}$ in $(\ref{auxiliary})$ are not independent
but are related via the duality condition $(\ref{constraintlow})$.
Note that if   this constraint were applied  to the action, it
would make the gauge field kinetic term $(\ref{quarter})$ vanish.
This was to be expected as the twisted self-duality condition
$(\ref{constraint2})$, from which the duality condition
($\ref{constraintlow})$ is obtained, implies the vanishing of the
gauge kinetic term in the $D+1$-dimensional doubled lagrangian
$(\ref{doubled2})$. Thus it is important that one first varies the
action  and then imposes the constraint $(\ref{constraintlow})$.

It is straightforward to verify that the field equations derived
from $\CL_{b}$ for the potentials $ A_{n-1}$ are consistent with
the $D$ dimensional constraint $(\ref{constraintlow})$. After some
computation one finds that the condition for consistency is that
the mass matrix $M$ should satisfy the equation $(\ref{inf})$.

\subsection{Lagrangian for Reduced Theory}

The odd dimensional massive self-duality condition
(\ref{constraint3}) can be obtained from a Chern-Simons  action of
the form $(\ref{CS})$, as we now show. In the case in which $M$ is
invertible,  the $D$-dimensional constraint $(\ref{constraint3})$
follows from the following lagrangian:
\begin{equation}\label{true}
\CL_{b}' = \frac{1}{2}P_{ij} [(-1)^{n-1} A_{n-1}^{i} \wedge
DA_{n-1}^{j} +  e^{\gamma} \tilde{M}^{j}_{\ k} A_{n-1}^{i} \wedge
* A_{n-1}^{k}],
\end{equation}
where $P_{ij}$ is any invertible matrix satisfying $P^T= (-1)^nP$.
This generalises the lagrangian of \cite{Townsend:xs}.
    We now show that for the special choice
$P_{ij}=(\Omega^{-1} M)_{ij}$, varying this action with respect to
the scalars, metric and other fields also give the right equations
of motion. Note that it follows from $(\ref{inf})$ that $P =
\Omega^{-1} M$ is a symmetric matrix if $n$ is even and is
antisymmetric if $n$ is odd. Similarly, $P \tilde{M}$ is a
symmetric matrix.

Now consider the lagrangian
\begin{equation}\label{lagrangian2}
\CL_{D}' = \CL_{g} + \CL_{s} + \CL_{b}'
\end{equation}
\no where $\CL_{g}$ and $\CL_{s}$ are as in $(\ref{auxiliary})$.
We will show that  $\CL_{D}'$ is equivalent to the lagrangian
$\CL_{D}$ (\ref{lagrangian}) with $ \CL_{b}$ given by
(\ref{quarter}) in the sense that they yield the same field
equations for all fields when the field equations of $\CL_{D}$ are
supplemented by the $D$-dimensional constraint; the analysis is
similar to that in \cite{julia}.

The field equations for the potential fields $A_{n-1}$ have
already been discussed. Now we check  the field equations for the
scalar fields. Let $\kappa$ represent any of the  scalar fields in
the theory except for the Kaluza-Klein field $\varphi$. Then
\begin{eqnarray}\label{check}
\delta_{\kappa} \CL_{b} & = & -\frac{1}{4} e^{-\gamma} H_{n}^{T}
\frac{\delta \CK}{\delta \kappa}\wedge * H_{n} - \frac{1}{4}
e^{\gamma} H_{n-1}^{T} \frac{\delta
\CK}{\delta \kappa}\wedge * H_{n-1} \nonumber \\
& = & - \frac{e^{\gamma}}{4} A_{n-1}^{T} \tilde{M}^{T}
\frac{\delta \CK}{\delta \kappa} \tilde{M} \wedge * A_{n-1} -
\frac{e^{\gamma}}{4} A_{n-1}^{T} M^{T}
\frac{\delta \CK}{\delta \kappa} M \wedge * A_{n-1} \nonumber \\
& = & - \frac{e^{\gamma}}{2} A_{n-1}^{T} M^{T} \frac{\delta
\CK}{\delta \kappa} M \wedge * A_{n-1} = \delta_{\kappa} \CL_{b}'.
\end{eqnarray}
\noindent In the second line we have imposed the constraint
$(\ref{constraint3})$. In the third line we used the symmetry
properties of the matrices $\Omega$ and $\CK$, the fact that
$\Omega$ is $G$-invariant and also that $\frac{\delta \CK}{\delta
\kappa} \ \Omega \CK = - \CK \Omega \ \frac{\delta \CK}{\delta
\kappa}$. The last equality in $(\ref{check})$ holds because $P
\tilde{M} = -M^T \Omega^{-1} \Omega \CK M = -M^{T} \CK M$. This
establishes that the two lagrangians $\CL$ and $\CL'$ have the
same field equations  for the scalar fields.

In order to check the equivalence of the field equations for the
metric, it is useful to note the following relation:
\begin{equation}\label{maybe1}
\frac{\delta}{\delta g^{\alpha \beta}}(H_{n}^{T} \CK \wedge *
H_{n}) = \CK_{ij} \sqrt{-g} (-n H_{n}^{(i) \alpha \mu_{1} \cdots
\mu_{n-1}} \ H^{(j) \beta}_{n \ \ \ \mu_{1} \cdots \mu_{n-1}}+
\frac{1}{2} g^{\alpha \beta} H_{n}^{(i) \mu_{1} \cdots \mu_{n}}
H^{(j)}_{n \ \mu_{1} \cdots \mu_{n}}).
\end{equation}
    Two such terms come from the variation of $\CL_{b}$ in
$(\ref{quarter})$. Imposing the constraint
$(\ref{constraintlow})$ on these terms and then using the
properties of the matrices $\CK$, $M$ and $\Omega$ as before one
can show that
\begin{eqnarray}\label{maybe2}
\frac{\delta \CL_{b}}{\delta g^{\alpha \beta}}& = &
\frac{e^{\gamma}}{2(n-1)!} \sqrt{-g} \bar{M}_{kl} \left((n-1)
A_{(n-1)}^{(k) \alpha \sigma_{1} \cdots \sigma_{n-2}} A^{(l)
\beta}_{(n-1) \sigma_{1} \cdots \sigma_{n-2}}
     -\frac{1}{2} g^{\alpha \beta} A_{(n-1) \sigma_{1} \cdots
\sigma_{n-1}}^{(k)} A_{(n-1)}^{(l) \sigma_{1} \cdots \sigma_{n-1}}
\right) \nonumber \\
&=& -\frac{1}{2}e^{\gamma} \frac{\delta}{\delta g^{\alpha
\beta}}(\bar{M}_{kl} A^{(k)}_{n-1} \wedge * A^{(l)}_{n-1}) =
\frac{\delta \CL_{b}'}{\delta g^{\alpha \beta}}
\end{eqnarray}
where we have defined $\bar{M}_{kl} = \CK_{ij} M^{i}_{\ k}
M^{j}_{\ l} = (M^{T} \CK M)_{kl}$.   The equivalence of the field
equations for the Kaluza-Klein field $\varphi$ are also easily
checked.

As a result we have a new D-dimensional lagrangian which yields
the D-dimensional field equations  and also the constraint:
\begin{eqnarray}
\CL_{D} & = &  R * 1 - \frac{1}{2} d\varphi \wedge * d\varphi -
\frac{1}{2} e^{-2(D-1) \alpha \varphi}
\CF_{2} \wedge * \CF_{2}  \label{new} \\
      &  & + \frac{1}{2}(\Omega^{-1} M)_{ij} [(-1)^{n-1}A_{n-1}^{i}
\wedge
DA_{n-1}^{j} +  e^{\gamma} \tilde{M}^{j}_{\ k} A_{n-1}^{i} \wedge
* A_{n-1}^{k}] \nonumber
\\  &  & + \frac{1}{4} {\rm
tr}(\CD\CK \wedge * \CD\CK^{-1}) - \frac{1}{2} e^{2(D-1) \alpha
\varphi} {\rm tr}(M^{2} + M \CK^{-1} M^{T} \CK) * 1. \nonumber
\end{eqnarray}

If $M$ is not invertible, there is a similar action with a
Chern-Simons action for the massive $n-1$ form gauge fields, and a
standard action for the massless gauge fields. First note that the
lagrangian (\ref{true}) can be written in flat indices as
\begin{equation}\label{deneme42}
    \CL_{b}' = \frac{1}{2}P_{ab} [(-1)^{n-1} A_{n-1}^{a} \wedge
\tilde \CD A_{n-1}^{b} +  e^{\gamma} \tilde{M}^{b}_{\ c}
A_{n-1}^{a} \wedge * A_{n-1}^{c}],
\end{equation}
where $P_{ab} = P_{ij} (\CV^{-1})^{i}_{\ a} (\CV^{-1})^{j}_{\ b} =
(\Omega^{-1})_{ac} M^{c}_{\ \ b} $ and $\tilde{M}^{a}_{\ \ b} =
\tilde{M}^{i}_{\ j} \CV_{\ i}^{a} (\CV^{-1})^{j}_{\ b}$. Note that
one has $P^{ab} = P_{cd} \Omega^{ca} \Omega^{db} = (-1)^{n-1}
M^{ab}$.
When $M$ is not invertible, $A_{n-1}^\alpha$ drops out from this
lagrangian, which is now just a lagrangian for
$\bar{A}_{(n-1)\alpha '}$:
\begin{equation}\label{deneme6.5}
\CL_{b1}' = \frac{1}{2}m^{\alpha' \beta'}[\bar{A}_{(n-1) \alpha'}
\wedge \tilde  \CD \bar{A}_{(n-1) \beta'} +(-1)^{n-1} e^{\gamma}
\delta_{\beta' \gamma'} m^{\gamma' \rho'} \bar{A}_{(n-1) \alpha'}
\wedge * \bar{A}_{(n-1) \rho'}].
\end{equation}
Here we have used that $\tilde{M} = QM = \Omega \CK M$ so that
$\tilde{M}^{a}{}_{b} = \Omega^{ac} \delta_{cd} M^{d}{}_{b}$.  Note
that $m^{\alpha' \beta'} = (-1)^n m^{\beta' \alpha'}$ because of
(\ref{M2}) and $m^{\alpha' \beta'} \delta_{\beta' \gamma'}
m^{\gamma' \rho'}$ is always symmetric, as it should be. It is
easy to see that the field equations of (\ref{deneme6.5}) for the
gauge fields $\bar{A}_{(n-1) \alpha'}$ does indeed give the
constraint (\ref{deneme5}).
    The lagrangian for $A^\alpha $   arises from
$(\ref{deneme3})$ (with an extra factor of $1/2$):
\begin{equation}\label{deneme7}
\CL_{b 2}  =  -\frac{1}{4}e^{-\gamma} \delta^{\alpha \beta}
\bar{H}_{(n) \alpha} \wedge * \bar{H}_{(n) \beta}- \frac{1}{4}
e^{\gamma} \delta_{\alpha \beta} \CD A_{n-2}^{\alpha}\wedge * \CD
A_{n-2}^{\beta}
\end{equation}
    subject to the constraint (\ref{deneme6}), which can be used to
eliminate either $A^\alpha _{n-2}$ or $A^\alpha _{n-1}$. Choosing
the first, the lagrangian for $A^\alpha _{n-1}$ is
\begin{equation}\label{deneme99}
\CL_{b 2}'  =
     - \frac{1}{2} e^{\gamma} \delta^{\alpha \beta}
\CD \bar{A}_{(n-1) \alpha} \wedge * \CD \bar{A}_{(n-1) \beta}.
\end{equation}
   Then the total lagrangian is
\begin{equation}\label{newlag}
      \CL_{D}' = \CL_{g} + \CL_{s} + \CL_{b1}' + \CL_{b 2}'
\end{equation}
where $\CL_{g}$ and $\CL_{s}$ are as in (\ref{auxiliary}).
   It is straightforward
to show that these give the right field equations, by an argument
similar to that in the invertible case above.

\subsection{G = SL(2, \IR) Case}

In this subsection we will consider the case $G = SL(2, \IR)$. In
this case the matrices $\CK$ and $\Omega$ are as in
$(\ref{matrix})$ and $(\ref{Omega})$. There are three distinct
reductions corresponding to the three conjugacy classes of $SL(2,
\IR)$ as discussed in section 2. The mass matrices representing
the three conjugacy classes are given in $(\ref{sl2mass})$. Now we
will give the reduced lagrangians for each mass matrix $M_{e}$,
$M_{h}$ and $M_{p}$.

\bigskip

$M_{e}$:

There are two massive, $(n-1)$-forms in the theory which we will
call $A^{1}$ and $A^{2}$. This is an $SO(2)$-gauged theory since
$M_{e}$ generates the $SO(2)$ subgroup of $SL(2, \IR)$.
(If $n=2$, there are additional gauge fields and the gauge group is
$ISO(2)$.)
This is
the only case the theory has a stable minimum of the potential
$\cite{atish}$. The global minimum of the potential is at $\chi
=\phi=0$. The lagrangian is:

\begin{eqnarray}
\CL_{D} & = &  R * 1 - \frac{1}{2} d\varphi \wedge * d\varphi -
\frac{1}{2} e^{-2(D-1) \alpha \varphi}
\CF_{2} \wedge * \CF_{2}  \label{elliptic} \\
      &  & + \frac{1}{2}m \{(-1)^{n-1}A^{1} \wedge DA^{1} +(-1)^{n-1}
A^{2} \wedge DA^{2} - m e^{\gamma} e^{\phi}[A^{1} \wedge *
A^{1} \nonumber \\
& &+ (e^{-2 \phi} + \chi^{2})A^{2} \wedge * A^{2}+2 \chi A^{1}
\wedge * A^{2}]\} \nonumber
\\  &  & + \frac{1}{4} {\rm
tr}(\CD\CK \wedge * \CD\CK^{-1}) - 2 e^{2(D-1) \alpha \varphi}
m^{2}[\sinh^{2}\phi + \chi^{2}(2+e^{2 \phi}(2+\chi^{2}))] * 1.
\nonumber
\end{eqnarray}

\bigskip

$M_{h}$:

There are two massive, $(n-1)$-forms in the theory which we will
call $A^{1}$ and $A^{2}$, as before. The gauge group is $SO(1,1)$
in this case (for $n>2$). The lagrangian is:


\begin{eqnarray}
\CL_{D} & = &  R * 1 - \frac{1}{2} d\varphi \wedge * d\varphi -
\frac{1}{2} e^{-2(D-1) \alpha \varphi}
\CF_{2} \wedge * \CF_{2}  \label{hyperbolic} \\
      &  & + \frac{1}{2}m \{(-1)^{n-1}2A^{1} \wedge DA^{2}
- m e^{\gamma} e^{\phi}[\chi A^{1} \wedge *
A^{1} \nonumber \\
& &+ \chi A^{2} \wedge * A^{2} +(e^{-2 \phi} + \chi^{2}+1)A^{1}
\wedge * A^{2}]\} \nonumber
\\  &  & + \frac{1}{4} {\rm
tr}(\CD\CK \wedge * \CD\CK^{-1}) -  2 e^{2(D-1) \alpha \varphi}
m^{2}[ 1 + \chi^{2}e^{2 \phi}] * 1. \nonumber
\end{eqnarray}

\bigskip

$M_{p}$:

There is one massive $(n-1)$-form field $\bar{A}_{1}$, one
massless $(n-1)$-form field $\bar{A}_{2}$ and one massless
$(n-2)$-form field $B^{2}$. However one can eliminate $B^{2}$ by
using the reduced constraint $(\ref{deneme6})$, as was discussed
in the previous subsection. The gauge group is $SO(1,1)$ in this
case (for $n>2$).
\begin{eqnarray}
\CL_{D} & = &  R * 1 - \frac{1}{2} d\varphi \wedge * d\varphi -
\frac{1}{2} e^{-2(D-1) \alpha \varphi}
\CF_{2} \wedge * \CF_{2}  \label{parabolic} \\
      &  & +\frac{1}{2}m[\bar{A}_{1} \wedge \CD \bar{A}_{1} +
(-1)^{n-1}
      e^{\gamma} m \bar{A}_{1} \wedge * \bar{A}_{1}] -
      \frac{1}{2} e^{\gamma} \CD \bar{A}_{2} \wedge *
      \CD \bar{A}_{2} \nonumber
\\  &  & + \frac{1}{4} {\rm
tr}(\CD\CK \wedge * \CD\CK^{-1}) -  \frac{1}{2} e^{2(D-1) \alpha
\varphi} m^{2}(e^{-\phi} + e^{\phi} \chi^{2})^{2} * 1. \nonumber
\end{eqnarray}

\section{Supergravity Applications}

In this section, we will apply our results to the twisted
reduction of supergravity theories in $d=D+1=4,6,8$ dimensions to
$D=3,5,7$. We will discuss general features here, and give details
of the full lagrangians  and of the classification of theories
elsewhere.

\subsection{Reduction of d=8 Maximal Supergravity}

The $N=2 \ d=8$ maximal supergravity $\cite{salam}$   can be
obtained from 11-dimensional supergravity by toroidal
compactification and has field equations invariant under the
duality group $SL(2, \IR) \times SL(3, \IR)$. The bosonic fields
consist of a metric, a 3-form gauge field $A_3$, 6 vector fields
in the {\bf (2,3)} representation of $SL(2, \IR) \times SL(3,
\IR)$, 3 2-form gauge fields in the {\bf (1,3)} representation of
$SL(2, \IR) \times SL(3, \IR)$, and scalars taking values in the
coset space $SL(3, \IR) / SO(3)\times SL(2,\IR) / SO(2)$. The
gauge field $A_3$ combines with the dual gauge field $\tilde A_3$
to form a doublet under  $SL(2, \IR) $ and  $SL(3, \IR)$   is a
symmetry of the action whereas $SL(2, \IR)$ is  a symmetry of the
field equations only, as it acts through electro-magnetic duality
on the 3-form gauge fields.

There is a consistent truncation of this theory where only the
$SL(3, \IR)$ singlets are kept and all the other fields are set to
zero $\cite{Izquierdo:1995ms}$. Then the truncated theory consists
of a metric, a 3-form gauge field and scalars taking values in
$SL(2,\IR) / SO(2)$, with an $SL(2, \IR)$ S-duality symmetry. This
truncated theory is precisely of the form ($\ref{lag1}$) with
$n=4$ and the twisted reduction with an $SL(2,\R)$ twist gives
three distinct reduced theories corresponding to the three
conjugacy classes, with lagrangians ($\ref{elliptic}$),
($\ref{hyperbolic}$) or ($\ref{parabolic}$).

This can be extended to the full theory, as the reduction of the
fields that are not $SL(3, \IR)$ singlets is a standard
Scherk-Schwarz reduction. There are some complications resulting
from the Chern-Simons interactions of the $d=8$ theory, and we
will not present the full results here. There are three distinct
classical theories, while the
distinct quantum theories correspond to the distinct $SL(2,\Z)$
conjugacy classes.

\subsection{Reduction of d=4, N=4 Supergravity}
   $N=4$ supergravity coupled to $p$ vector
multiplets has an $O(6,p) $ symmetry of the action and an
$SL(2,\IR)$ S-duality symmetry of the equations of motion. The
vector fields $A_1^I$ ($I=1,2,..., 6+p$) are in the fundamental
{\bf 6+p} representation of $O(6,p) $ and combine with dual
potentials $\tilde A_1^I$ to form $6+p$ doublets $A_1^{mI}$
($m=1,2$) transforming in the {\bf (2,6+p)} of  $SL(2,\IR)\times
O(6,p) $. The scalars take values in the coset
$SL(2,\IR)/SO(2)\times O(6, 22) / O(6) \times O(22)$. The scalars
in $O(6, 22) / O(6) \times O(22)$ can be represented by a coset space
metric $
\CN_{IJ}$ while the 2 scalars $\phi,\chi$  in
$SL(2,\IR)/SO(2)$ can be represented by a  coset
space metric   $\CK _{mn}$ which is of the same form as
($\ref{matrix}$).

The lagrangian for the bosonic sector can be written as
     $\cite{Sen:1994fa,Bergshoeff:1985ms,deRoo}$:
\begin{eqnarray}\label{lag5}
        \CL & = & R * 1 + \frac{1}{4} {\rm tr}(d\CK \wedge *
     d\CK^{-1}) + \frac{1}{4} {\rm tr}(d\CN
       \wedge * d\CN^{-1})- \frac{1}{2} e^{-\phi} F_2^{I}
\CN_{IJ}\wedge *
F_2^{J} \nonumber \\
       & & - \frac{1}{2} \chi F_2^{I} L_{IJ}\wedge F_2^{J}
\end{eqnarray}
where $L$ is the $O(6, p)$ invariant metric and the matrices $\CN$
and $L$ satisfy
\begin{equation}\label{NandL}
        \CN^{T} = \CN, \ \ \ \ \ \ \CN^{T} L \CN = L.
\end{equation}

Now the vector field equation can be written as $dG^I_2=0$ where
\begin{equation}\label{GGG}
G_{2}^{I} =  (L^{-1})^{IJ} \frac{\delta \CL}{\delta F_2^{J}} =
-e^{-\phi} \CR^{I}_{\ J} * F_2^{J} - \chi F_2^{I}
\end{equation}
and the matrix $\CR$ is defined as
\begin{equation}\label{R}
L_{PI} \CR^{I}_{\ J}  = \CN_{PJ}.
\end{equation}
Note that $\CR^{2} = 1$. Now we can write
\begin{eqnarray}
\CL' & = &  R * 1 + \frac{1}{4} {\rm tr}(d \CK \wedge * d
\CK^{-1}) + \frac{1}{4}  {\rm tr}(d \CN \wedge * d \CN^{-1})
\\ \label{lag4} &  & +  \frac{1}{2} F_2^{I} L_{IJ} \wedge G_2^{J}
\nonumber
\end{eqnarray}
As before, the field equations $dG_2^I=0$ imply the existence of
dual potentials $\tilde A_1^I$, with $G_{2}^{I}=d\tilde
A_{1}^{I}$. Then the full set of vector fields $A_1^{i}$ in the
doubled formalism
      is $A_1^{mI}= (A_1^I,\tilde A_1^I)$ where $i=1,..., 2(6+p)$
becomes
the composite index $mI$.
      The field strengths are the $6+p$
       $SL(2, \IR)$-doublets:
\begin{equation}\label{H}
        H_2^{I} = \left(\begin{array}{c}
                         dA_1^{I} \\
                         d\tilde{A}_1^{I}
                         \end{array}\right).
\end{equation}

     \no We also impose the twisted self-duality constraint
\begin{equation}\label{constraint5}
H_2^{mI} = J^{m}{}_{ n} \CR^{I}_{\ J} * H_2^{nJ}.
\end{equation}
where $J^{m}{}_{ n} $ is as in ($\ref{J}$), $J^{m}{}_{ n} =\Omega
^{mp} \CK _{pn}$. So the matrix $Q$ in $(\ref{constraint2})$ is
now the $(12+2p) \times (12+2p)$ matrix
\begin{equation}\label{56}
        Q = J \otimes \CR
\end{equation}
which satisfies $Q^{2}= -1$ since $J^{2} = -1 $ and $R^{2} = +1$.
The doubled lagrangian
\begin{eqnarray}
\CL & = & R * 1 + \frac{1}{4}  {\rm tr}(d \CK \wedge * d \CK^{-1})
+ \frac{1}{4}  {\rm tr}(d \CN \wedge * d \CN^{-1})
\label{doubled3}
\\ & & - \frac{1}{4}\CN_{IJ} H_2^{mI} \CK_{mn}\wedge * H_2^{nJ}.
\nonumber
\end{eqnarray}
gives the same field equations as those of $(\ref{lag5})$ when the
constraint equation $(\ref{constraint5})$ is imposed \cite{julia}.
This lagrangian is of the same form as $(\ref{doubled})$, with
$\CK_{ij}$ given by
$$
\CK_{mI~ nJ}=\CK_{mn}\CN_{IJ}
$$
Then the Scherk-Schwarz reduction of $(\ref{doubled3}$) with
mass matrix $M^m{}_n$ in the Lie algebra of $SL(2,\R)$ can be
performed as before and the three dimensional lagrangian that one
obtains is:
\begin{eqnarray}
\CL_{3}' & = &  R * 1 - \frac{1}{2} d\varphi \wedge * d\varphi -
\frac{1}{2} e^{- 2\varphi} \CF_{2} \wedge * \CF_{2} + \frac{1}{8}
{\rm tr}(d \CN \wedge
* d \CN^{-1}) \label{three1} \\
& - & \frac{1}{4} e^{- \varphi} \CN_{IJ} H^{mI}_{2} \CK_{mn}
\wedge * H_{n2}^{J} - \frac{1}{4} e^{\varphi} \CN_{IJ} H^{mI}_{1}
\CK_{mn}
\wedge * H_{1}^{nJ} \nonumber \\
     & +  & \frac{1}{4} {\rm
tr}(\CD\CK \wedge * \CD\CK^{-1}) - \frac{1}{2} e^{2 \varphi} {\rm
tr}(M^{2} + M \CK^{-1} M^{T} \CK) * 1. \nonumber
\end{eqnarray}
This lagrangian is to be supplemented by the reduced constraint
\begin{equation}\label{concon}
H_2^{mI} = e^{\varphi} J^{m}{}_{ n} \CR^{I}_{\ J} * H_1^{nJ}.
\end{equation}
When $M$ is invertible, this becomes
\begin{equation}\label{constraint7}
DA_1^{mI} = e^{\varphi} J^{m}{}_{ n} \CR^{I}_{\ J} M^{n}_{\ p}*
A_1^{pJ},
\end{equation}
after gauging the Stuckelberg fields away, as in section 2.
     As before one can find a  three dimensional lagrangian from which
the field equations and the constraint can be derived. This
lagrangian is (for invertible $M$):
\begin{eqnarray}
\CL_{3} & = &  R * 1 - \frac{1}{2} d\varphi \wedge * d\varphi -
\frac{1}{2} e^{-2 \varphi} \CF_{2} \wedge * \CF_{2} + \frac{1}{4}
{\rm tr}(d \CN \wedge * d \CN^{-1}) \label{three2}
\\ & + & \frac{1}{2}(\Omega^{-1} M)_{mn} (-L_{IJ} A_{1}^{mI} \wedge
DA_{1}^{nJ}
     + \CN_{IJ} e^{\varphi}
\tilde{M}^{n}_{\ p} A_{1}^{mI} \wedge * A_{1}^{pJ})
     \nonumber
\\ & +  & \frac{1}{4} {\rm
tr}(\CD\CK \wedge * \CD\CK^{-1}) - \frac{1}{2} e^{2 \varphi} {\rm
tr}(M^{2} + M \CK^{-1} M^{T} \CK) * 1. \nonumber
\end{eqnarray}
There is a similar action for the case in which $M$ is non-invertible.

\subsection{Reduction of d=4,N=8 Supergravity}

The $D=4,N=8$ theory has $E_7$ duality symmetry of the equations
of motion. There are 70 scalars taking values in the coset
$E_7/SU(8)$, and 28 vector fields $A^I$ which combine with their
duals to give $A^i$ transforming as a {\bf 56} of $E_7$. The
bosonic action can be written as ($\ref{doubled}$) with the
constraint ($\ref{constraint2}$) where $Q$ is as in $(\ref{Q})$
and $\Omega^{ij}$ is the symplectic invariant of $E_7$
$\cite{cremmerjulia}$. Now $\CK$ is the matrix which parametrizes
the scalar coset $E_{7}/SU(8)$. The theory can be reduced to
3-dimensions using any mass matrix $M$ in the Lie algebra of
$E_7$. Naively, this introduces 133 mass parameters, but these
theories are not all independent and the independent theories
correspond to the distinct conjugacy classes;  the
classification of conjugacy classes in this case is not known. The
matrix
$M^{ab} = M^{a}_{\ c} \Omega^{cb}$   introduced in section 2  is a
symmetric matrix since $n=2$ in $(\ref{M2})$ so, by choosing   a
suitable basis, it can be brought into the diagonal form:
\begin{equation}\label{e7}
       M^{ab} = \left(\begin{array}{ccc}
                          m_{1} & & \bigcirc \\
                          & \ddots & \\
                          \bigcirc & & m_{56}
                          \end{array}\right)
\end{equation}
For example, consider
      performing
the Scherk-Schwarz reduction with   the Lie algebra element
$M^{ab}$ of the form:
\begin{equation}\label{e7mass}
       M^{ab} = m \left(\begin{array}{ccc}
                          0_{l} & & \bigcirc \\
                          & 1_{p}& \\
                          \bigcirc & & -1_{q}
                          \end{array}\right)
\end{equation}
where $l+p+q = 56$. Then one obtains a 3-dimensional theory with
one mass parameter with $p $ massive, self-dual vector fields, $q
$ massive, anti-self-dual vector fields
    and $l$ massless vector fields which are dual to the $l$
massless scalar fields coming from the reduction of the vector
field in the 4-dimensional theory.

\subsection{Reduction of d=6 Supergravity}
   The $d=6$ theory of
$\cite{ortin}$, obtained from a  truncation of the toroidal
compactification
      of IIB supergravity,
      has an
\begin{equation}\label{symmetry}
SO(2, 2) \equiv SL(2, \IR)_{EM} \times SL(2, \IR)_{IIB}
\end{equation}
\no symmetry of the equations of motion. The  $SL(2,\IR)_{IIB}$ is
inherited from the $SL(2, \IR)$ symmetry of IIB in ten dimensions
and   is   a symmetry of the action in the six dimensional theory.
However $SL(2, \IR)_{EM}$ is a symmetry of the field equations
only. The bosonic lagrangian is:
\begin{eqnarray}\label{six}
\CL & = & R * 1 + \frac{1}{4} {\rm tr}(d \CK \wedge * d \CK^{-1})
+ \frac{1}{4} {\rm tr}(d \CN \wedge * d\CN^{-1})
\nonumber \\
&& - \frac{1}{2}e^{-\phi_{1}} F^{I} \CN_{IJ}\wedge * F^{J} -
\frac{1}{2}\chi_{1} F^{I} \Omega_{IJ} \wedge F^{J}
\end{eqnarray}
Here $F^{I} = dA_{2}^{I}$ are the two 3-form field strengths and
$I, J = 1, 2$ are $SL(2, \IR)_{IIB}$ indices. We also introduce
$SL(2, \IR)_{EM}$ indices $m,n=1,2$. There are two $SL(2,
\IR)/SO(2)$ scalar cosets in the theory. $e^{-\phi_{1}}$ and
$\chi_{1}$ parametrize the scalar coset $SL(2, \IR)_{EM}/SO(2)$,
represented by the matrix $\CK_{mn}$. The other two scalars
$e^{-\phi_{2}}$ and $\chi_{2}$, parametrize the scalar coset
$SL(2, \IR)_{IIB}/SO(2)$, which is represented by the matrix
$\CN_{IJ}$. The invariant matrices are $\Omega^{mn}$,
$\Omega^{IJ}$.

The lagrangian $(\ref{six})$ is of the same form as
$(\ref{lag5})$, where now $I$ ranges from 1 to 2 and the $O(6, p)$
invariant $L_{IJ}$ has been replaced by the $SL(2, \R)_{IIB}$
invariant matrix $\Omega_{IJ}$. So $(\ref{six})$ is equivalent to
the doubled lagrangian $(\ref{doubled3})$ (now with 3-form field
strengths $H_{3}$) when supplemented by the constraint
$(\ref{constraint5})$. Note that the matrix $Q$ in ($\ref{56})$
now satisfies $Q^{2} = +1$, as it should in 6 dimensions, since
now $\CR^{2} = -1$, whereas $\CR^{2} = +1$ and hence $Q^{2} = -1$
in the 4-dimensional case.

By performing the Scherk-Schwarz reduction of the doubled
lagrangian with monodromy in $SL(2, \IR)_{EM}$,
one obtains the following  auxiliary five-dimensional
lagrangian:
\begin{eqnarray}
\CL_{5}' & = &  R * 1 - \frac{1}{2} d\varphi \wedge * d\varphi -
\frac{1}{2} e^{-4/\sqrt{6} \varphi} \CF_{2} \wedge * \CF_{2} +
\frac{1}{4} {\rm tr}(d \CN \wedge
* d \CN^{-1}) \label{five1} \\
& - & \frac{1}{4} e^{-2/\sqrt{6} \varphi} \CN_{IJ} H^{mI}_{3}
\CK_{mn} \wedge * H_{3}^{nJ} - \frac{1}{4} e^{2/\sqrt{6} \varphi}
\CN_{IJ} H^{mI}_{2} \CK_{mn}
\wedge * H_{2}^{nJ} \nonumber \\
     & +  & \frac{1}{4} {\rm
tr}(\CD\CK \wedge * \CD\CK^{-1}) - \frac{1}{2} e^{4/\sqrt{6}
\varphi} {\rm tr}(M^{2} + M \CK^{-1} M^{T} \CK) * 1 \nonumber
\end{eqnarray}
This is to be supplemented by the five-dimensional reduced
constraint
\begin{equation}\label{concon2}
H_3^{mI} = e^{2/\sqrt{6} \varphi} J^{m}{}_{ n} \CR^{I}_{\ J} *
H_2^{nJ}.
\end{equation}
When $M$ is invertible one can gauge the Stuckelberg fields away
and in this gauge  the constraint in (\ref{concon2}) takes the
form
\begin{equation}\label{constraint8}
DA_2^{mI} = - e^{2/\sqrt{6} \varphi} J^{m}{}_{ n} \CR^{I}_{\ J}
M^{n}_{\ p}* A_2^{pJ}.
\end{equation}

The five dimensional reduced lagrangian  from which the reduced
constraint $(\ref{constraint8})$   and the field equations of
$(\ref{five1}$) can be derived is obtained by using the techniques
of the previous sections:
\begin{eqnarray}
\CL_{5} & = &  R * 1 - \frac{1}{2} d\varphi \wedge * d\varphi -
\frac{1}{2} e^{-4/\sqrt{6} \varphi} \CF_{2} \wedge * \CF_{2} +
\frac{1}{4} {\rm tr}(d \CN \wedge * d \CN^{-1}) \label{five2}
\\ & + & \frac{1}{2}(\Omega^{-1}M)_{mn} (\Omega_{IJ} A_{2}^{mI} \wedge
DA_{2}^{nJ}
     + \CN_{IJ} e^{2/\sqrt{6} \varphi}
\tilde{M}^{n}_{\ p} A_{2}^{nI} \wedge * A_{2}^{pJ})
     \nonumber
\\ & +  & \frac{1}{4} {\rm
tr}(\CD\CK \wedge * \CD\CK^{-1}) - \frac{1}{2} e^{4/\sqrt{6}
\varphi} {\rm tr}(M^{2} + M \CK^{-1} M^{T} \CK) * 1. \nonumber
\end{eqnarray}

\subsection{Reduction of d=6, N=8 Supergravity}
The maximal supergravity in six dimensions
     has  noncompact global symmetry
$SO(5,5)$, which can be realized at the level of field equations
only $\cite{tanii}$. There are five 3-form field strengths which
split into five self-dual ones and five anti-self dual ones, and
these ten transform as a {\bf 10} of $SO(5,5)$. There are 25
scalar fields in the theory and they parametrize the coset space
$SO(5,5)/SO(5) \times SO(5)$. The bosonic lagrangian can be written as
($\ref{doubled}$), plus terms  which we will not give explicitly here
involving the vector fields,
  with the constraint ($\ref{constraint2}$) where
$Q$ is as in $(\ref{Q})$ and $\Omega^{ij}$ is the symplectic
invariant of $SO(5,5)$ $\cite{julia}$. Now $\CK$ is the matrix
which parametrizes the scalar coset $SO(5,5)/SO(5) \times SO(5)$.
The theory can be reduced to 5-dimensions using any mass matrix
$M$ in the Lie algebra of $SO(5,5)$. The number of distinct
reductions is given by the number of conjugacy classes of
$SO(5,5)$.

Consider the matrix $M^{ab} = M^{a}_{\ c} \Omega^{cb}$
introduced in section 2. It is an anti-symmetric matrix since
$n=3$ in $(\ref{M2})$. So in  a particular basis it can be brought
into the skew-diagonal form:
\begin{equation}\label{so55}
       M^{ab} = \left(\begin{array}{ccccc}
                       0 & m_{1} & & \bigcirc & \\
                       -m_{1} & 0 & & & \\
                       & &  \ddots & & \\
                        \bigcirc & & & 0 & m_{5} \\
                       &  & & -m_{5} & 0
                       \end{array}\right)
\end{equation}
Consider a mass matrix of the form
\begin{equation}\label{so55mass}
       M^{ab} = m \left(\begin{array}{ccc}
                        {\bf 0}_{l} & & \\
                       & & \begin{array}{rrrrrr}
                       0 & 1 & & & &   \\
                       -1 & 0 & & & & \\
                       & & & \ddots & & \\
                       & & &  & & \\
                        & &  & & 0 & 1 \\
                        & & & & -1 & 0
                       \end{array}
                       \end{array}\right)
\end{equation}
  where there are $l$ zero
eigenvalues and the number of skew-diagonal blocks is $(10-l)/2$.
On reduction, one obtains, in five dimensions, a gauged theory
with one mass parameter including $10-l$ massive self-dual 2-form
fields and $l$ massless 2-form fields, which could be dualised to
$l$ massless 1-form fields.


\section*{Acknowledgments}

A. \c{C}.-\"{O}. would like to thank Tekin Dereli for support and
discussions. The work of A. \c{C}.-\"{O} has been supported by
T\"{U}B\.{I}TAK (Scientific and Technical Research Council of
Turkey) through the BAYG-BDP program.



\end{document}